\documentclass[11pt,a4paper]{article}
\usepackage{color}
\usepackage{indentfirst}
\usepackage{amsmath}
\usepackage{amssymb}
\usepackage[utf8]{inputenc}
\usepackage{amsfonts}
\usepackage{academicons}
\usepackage[pdftex]{hyperref}
\usepackage{xcolor}
\usepackage{hyperref}
\usepackage{amsmath}
    \numberwithin{equation}{section}
    \numberwithin{table}{section}
    \numberwithin{figure}{section}
\usepackage[header, toc]{appendix}

\newcommand{\be}{\begin{equation}}
\newcommand{\ee}{\end{equation}}
\newcommand{\ba}{\begin{eqnarray}}
\newcommand{\ea}{\end{eqnarray}}

\newcommand{\orcid}[1]{\href{https://orcid.org/#1}{\textcolor[HTML]{A6CE39}{\aiOrcid}}}

\begin{document}

\begin{center}


{\Large \bf New Faraday lines through Four Bosons EM}\\	
{\large R. Doria$^{\dagger}$\footnote{doria@aprendanet.com} and}
{\large L.S. Mendes$^{\dagger}$ \footnote{santiago.petropolis@outlook.com}}\\[0.5cm]

{\large $^\dagger$Aprendanet, Petropolis, Brazil; Quarks, Petropolis, Brazil}\\

\end{center}







\date{\today}

\begin{abstract}

Field physics was founded by Faraday introducing magnetic fields (1831), electric fields (1837) and light as an EM wave (1846), initiating the process where nature is made by matter and fields. Consider that, ordinary space is full of fields. The Faraday view is basis for modern quantum field theory.

The concept of fields set up a physicality in development. Physics would like to know how far matter is created by fields. Generate matter from nonlinear fields. Faraday lines of force relating physical entities as electric charge and mass depending on fields. 

Our purpose is on Faraday lines for nonlinear abelian electromagnetism. Introduce the Four Bosons EM. The phenomenology of a generic charge $\{+,0,-\}$ transmitted by four bosons $\{A_{\mu}, U_{\mu}, V_{\mu}^{\pm}\}$. Nonlinear equations constituted. New Faraday lines were introduced. The potentials fields of physics are developed. Granular and collective fields strengths expressed. Four types of fields charges are derived. They are electric charge, modulated, neutral, Bianchi.

This work introduces a systematic procedure of associative physics. Mass and charge are generated due to the four fields interrelationships. Masses are derived without spontaneous symmetry breaking. It is obtained naturally from gauge symmetry, London, and mixing terms. Electric charge is written by fields through the Noether theorem. EM interactions not necessarily coupled with electric charge are proposed. An enlargement of EM energy is derived.

\end{abstract}


\section{Introduction}

Faraday pioneered the physics dependence on fields [1]. The outrageous proposal of invisible lines discovered that physics is more than space, time, and movement. It contains the fields of physicality. In 1911, Rutherford not only discovered the atom as an empty space [2]. The greek atm would contain nuclei, electrons, and field.  In 1913, Bohr's orbit theory postulated a relationship between fields and matter in order to justify atom stability [3]. The assumption that there is physics more than matter became visible.

History shows that there is no complete nothingness. The bussole's navigators were guided by magnetic fields. Astrophysics discovered that the universe is more than an empty space. It contains a large cosmic structure at $10^{17}km$ with weak magnetic fields [4]. A magnetar was detected with a strong magnetic field at order $10^{19}T$ [5]. The QED Sauter-Schwinger effect says that electron-positron pairs are spontaneously created in the presence of an electric field [6].

Physics is more than matter. There is a field of physics to be exploited. Since Faraday fields became the variables of physics equations. A quantum field theory is ruling that particles are local vibrations of fields. Providing a fields dynamics, with conservation laws and associated fields charges. A realism to be taken.

Thus, a Faraday research is to understand how far the relationship between matter and fields is still unknown in electromagnetic theory. Maxwell equations are the origin of fields events [7]. It provides fields under Gauss and Ampere dynamics, Faraday induction, and monopoles absence. However, Maxwell is incomplete. It restricts fields to dependence on electric charge, linearity, and subsidiary potential fields. On other hand, QED introduces the particles creation and annihilation. Electron absorb or emit a photon; electron and positron annihilate and create two photons. However, despite $E=mc^{2}$, where matter is transformed into energy, but also, energy turn into matter, it does not study charges transfer. The quantum mutation phenomena responsible for the matter-energy transformation is not being supported by a theory. 

An electric charge transfer model beyond Maxwell and QED is expected. A candidate is the so-called four bosons of electromagnetism [8]. Based on the transfer of three electric charge flavors, it defines the generic electric charge $\{+,0,-\}$. A charge transfer physics intermediate by four bosons $\{A_{\mu}, U_{\mu}, V_{\mu}^{\pm}\}$ is proposed. It provides $A_{\mu}$ as the usual photon, $U_{\mu}$ massive photon and $V_{\mu}^{\pm}$ charged photons. These fields are interrelated by a non-univoque electromagnetic symmetry [9].

A fundamental nonlinear four bosons electrodynamics is derived. New Faraday lines are generated. Vector potentials become a real physical quantity that conglomerates can be measured. Granular and collective fields strengths are expressed. Potentials fields are working as their own sources and interacting with granular and collective fields strengths. Based on non-linear abelian fields, the quadruplet derives its own charges. They are electric charge, modulated, neutral, and Bianchi. It yields a new relationship between mass, charges, and fields to be studied.
\section{Fields strengths}

The quadruplet $\{ A_{\mu}, U_{\mu}, V_{\mu}^{\pm}\}$ is associated by the electromagnetic symmetry. Different possibilities to consider their interrelations were studied [8-9]. Diverse EM physics conveyed by the quadruplet may be derived. They share to differ common Lagrangian and Noether theorem relationships, Lagrangian coefficients. 

Our choice for the quadruplet interrelationship is the following gauge transformation:
\begin{eqnarray}
     A{'}_{\mu} &=& A_{\mu} + k_{1}\partial_{\mu}\alpha
     \\
     U{'}_{\mu} &=& U_{\mu} + k_{2}\partial_{\mu}\alpha
     \\
     V^{+'}_{\mu} &=& e^{iq\alpha(x)}V^{+}_{\mu} 
     \\
     V^{-'}_{\mu} &=& e^{-iq\alpha(x)}V^{-}_{\mu} 
\end{eqnarray}

At vectorial expression, one gets

It yields the following granular antisymmetric fields strengths

\begin{eqnarray}
	F_{\mu \nu} &=   \partial_\mu A_\nu - \partial_{\nu}A_\mu , & F'_{\mu \nu} = F_{\mu \nu} 
	\\
	U_{\mu \nu} &=  \partial_\mu U_\nu - \partial_{\nu}U_\mu , & U'_{\mu \nu} = U_{\mu \nu}
	\\
	V_{\mu \nu}^\pm &=  D_\mu V_\nu^\pm - D_{\nu}V_\mu^\pm, &   V_{\mu \nu}^{\pm'} = e^{\pm i q \alpha } V_{\mu \nu}^\pm
\end{eqnarray}
where the covariant derivative is given by $D_{\mu} = \partial_{\mu} + i(q_{1}A_{\mu} + q_{2}U_{\mu})$. The couplings $q_{1}$ and $q_{2}$ are modulated electric charge given by $q_{1} = a\cdot q;$  $q_{2} = b\cdot q$, where $ak_{1}+bk_{2}=-1$.

Collective fields strengths are expressed as
\begin{eqnarray}
    \mathbf{e}^{[12]'}_{[\mu \nu]} &=& \frac{1}{2}\mathbf{e}_{[12]}\left(A_{\mu}U_{\mu} - U_{\mu}A_{\mu}\right) = \mathbf{e}^{[12]}_{[\mu \nu]}
    \\
    \mathbf{e}^{[+-]'}_{[\mu \nu]} &=& i\mathbf{e}_{[34]}\left(V^{+}_{\mu}V^{-}_{\nu} - V^{-}_{\mu}V^{+}_{\nu}\right) = \mathbf{e}^{[12]}_{[\mu \nu]}
\end{eqnarray}
and
\begin{eqnarray}
    \mathbf{e}^{[12+]}_{[\mu \nu]} &=& \frac{1}{2} \mathbf{e}_{[12]}\left[\left(A_{\mu}-U_{\mu}\right)V^{+}_{\nu} + \left(A_{\nu}-U_{\nu}\right)V^{+}_{\mu} \right]
    \\
    \mathbf{e}^{[12-]}_{[\mu \nu]} &=& \frac{1}{2} \mathbf{e}_{[12]}\left[\left(A_{\mu}-U_{\mu}\right)V^{-}_{\nu} + \left(A_{\nu}-U_{\nu}\right)V^{-}_{\mu} \right]
\end{eqnarray}
 transforming as
 \begin{equation}
  \mathbf{e}^{[12+]'}_{[\mu \nu]} = e^{iq\alpha(x)}\mathbf{e}^{[12+]}_{[\mu \nu]}\label{Campo coletivo 12+}, \end{equation}
 \begin{equation} 
  \mathbf{e}^{[12-]'} = e^{-iq\alpha(x)}\mathbf{e}^{[12-]}_{[\mu \nu]}\label{Campo coletivo 12-}
\end{equation}

For the symmetric sector:
\begin{equation}
	S_{\mu \nu 1} = \partial_\mu A_\nu + \partial_\nu A_\mu,	S_{\mu \nu 2} = \partial_\mu U_\nu + \partial_\nu U_\mu.
\end{equation}
the invariance is under the condition
\begin{equation}
    S^{1'}_{\mu \nu } + S^{2'}_{\mu \nu } = S^{1}_{\mu \nu } + S^{2}_{\mu \nu }
\end{equation}

Charged fields are transforming as
\begin{eqnarray}
S_{\mu \nu}^{\pm} &=  D_\mu V_\nu^{\pm} + D_\nu V_\mu^{\pm}, &  S_{\mu \nu }^{\pm'}=e^{\pm i q \alpha} S^{\pm}_{\mu \nu }
\end{eqnarray}

Longitudinal terms are written as
\begin{eqnarray}
	S_{\alpha 1 }^{\alpha} &= 2\partial_\alpha A^{\alpha},
	S_{\alpha 2 }^{\alpha} &= 2\partial_\alpha U^{\alpha}.
\end{eqnarray}
with
\begin{equation}
    S^{\alpha 1 '}_{\alpha} + S_{\alpha }^{\alpha 2'} = S^{\alpha 1 }_{\alpha} + S^{\alpha 2 }_{\alpha}
\end{equation} 	
and
\begin{eqnarray}
S_{\alpha}^{\alpha \pm} &=  D_\alpha V^{\alpha \pm}, &  S_{\alpha }^{\alpha \pm'}=e^{\pm i q \alpha} S_{\alpha}^{\alpha \pm}
\end{eqnarray}

The collective symmetric fields strengths are 
\begin{eqnarray}
    &&\mathbf{e}_{(\mu \nu)} = \mathbf{e}_{(11)}A_{\mu}A^{\nu} + \mathbf{e}_{(12)}\left(A_{\mu}U_{\nu} + U_{\mu}A_{\nu}\right) + \mathbf{e}_{(22)}U_{\mu}U_{\nu} = \mathbf{e}'_{(\mu \nu)}\nonumber
    \\
    &&\mathbf{e}^{(+-)}_{(\mu \nu)} =\mathbf{e}_{(34)}V^{+}_{\mu}V^{-}_{\nu} = \mathbf{e}^{(+-)'}_{(\mu \nu)}
\end{eqnarray}
and
\begin{eqnarray}
    &&\mathbf{e}^{\alpha}_{\alpha} = \mathbf{e}_{(11)}A_{\alpha}A^{\alpha} + 2\mathbf{e}_{(12)}A_{\alpha}U^{\alpha} + \mathbf{e}_{(22)}U_{\alpha}U^{\alpha} = \mathbf{e}^{\alpha'}_{\alpha}
    \\
    &&\mathbf{e}^{(+-)\alpha}_{\alpha} = \mathbf{e}_{(34)}V^{+}_{\alpha}V^{-\alpha} = \mathbf{e}^{(+-)\alpha'}_{\alpha}
\end{eqnarray}

Other collective fields strengths are included 
\begin{eqnarray}
    &&\mathbf{e}^{(12+)}_{(\mu \nu)} = \left[\mathbf{e}_{(11)}A_{\mu} + \mathbf{e}_{(12)}\left(A_{\mu} + U_{\mu}\right) + \mathbf{e}_{(22)}U_{\mu}\right]V^{+}_{\nu}
    \\
   && \mathbf{e}^{(12-)}_{(\mu \nu)} = \left[\mathbf{e}_{(11)}A_{\mu} + \mathbf{e}_{(12)}\left(A_{\mu} + U_{\mu}\right) + \mathbf{e}_{(22)}U_{\mu}\right]V^{-}_{\nu}
   \\
    &&\mathbf{e}^{(++)}_{(\mu \nu)}  = \frac{1}{2}\left(\mathbf{e}_{(33)}-\mathbf{e}_{(44)}\right)V_\mu^+V_\nu^+ 
    \\
    &&\mathbf{e}^{(--)}_{(\mu \nu)} = \frac{1}{2}\left(\mathbf{e}_{(33)}-\mathbf{e}_{(44)}\right)V_\mu^-V_\nu^-
\end{eqnarray}
transforming as

\begin{eqnarray}
  \mathbf{e}^{(12+)'}_{(\mu \nu)} &= e^{iq\alpha(x)}\mathbf{e}^{(12+)}_{(\mu \nu)},
  &\mathbf{e}^{(12-)'}_{(\mu \nu)} = e^{-iq\alpha}\mathbf{e}^{(12-)}_{(\mu \nu)}
  \\
  \mathbf{e}^{(++)'}_{(\mu \nu)} &= e^{2iq\alpha(x)}\mathbf{e}^{(++)'}_{(\mu \nu)},
  &\mathbf{e}^{(--)'}_{(\mu \nu)} = e^{-2iq\alpha(x)}\mathbf{e}^{(--)'}_{(\mu \nu)}
\end{eqnarray}

Similarly,
\begin{eqnarray}
    &&\mathbf{e}^{(12+)\alpha}_{\alpha} = \left[\mathbf{e}_{(11)}A_{\alpha} + \mathbf{e}_{(12)}\left(A_{\alpha} + U_{\alpha}\right) + \mathbf{e}_{(22)}U_{\alpha}\right]V^{+\alpha}, 
    \\
    &&\mathbf{e}^{(12-)\alpha}_{\alpha} = \left[\mathbf{e}_{(11)}A_{\alpha} + \mathbf{e}_{(12)}\left(A_{\alpha} + U_{\alpha}\right) + \mathbf{e}_{(22)}U_{\alpha}\right]V^{-\alpha}
    \\
    &&\mathbf{e}^{(++)\alpha}_{\alpha} = \frac{1}{2}\mathbf{e}_{(34)}V_{\alpha}^{+}V^{\alpha +}, \mathbf{e}^{(--)\alpha}_{\alpha} = \mathbf{e}_{(34)}V_{\alpha}^{-}V^{\alpha -}
\end{eqnarray}

\begin{eqnarray}
  &&\mathbf{e}^{(12+)\alpha}_{\alpha} = e^{iq_{2}\alpha(x)}\mathbf{e}^{(12+)\alpha}_{\alpha}, \mathbf{e}^{(12-)\alpha}_{\alpha} = e^{-iq_{2}\alpha(x)}\mathbf{e}^{(12-)\alpha}_{\alpha}
  \\
  &&\mathbf{e}^{(++)\alpha}_{\alpha} = e^{2 iq_{2}\alpha(x)} \mathbf{e}^{(++)\alpha}_{\alpha}, \mathbf{e}^{(--)\alpha}_{\alpha} = e^{-2iq_{2}\alpha} \mathbf{e}^{(--)\alpha}_{\alpha} 
\end{eqnarray}
The corresponding antisymmetric vectorial fields strengths are written as

\begin{align}
	F^I_{\mu \nu} &= \partial_{\mu} A^{I}_{\nu}-\partial_{\nu}A^{I}_{\mu};& \vec{E}_{i}^{I}&=F^{I}_{0i}; & \vec{B}_{i}^{I}&=\frac{1}{2}\epsilon_{ijk}F^{I}_{jk} 
\end{align}
where $A^{\mu}_{I} \equiv \left(\phi_{I}, \vec{A}_{I}\right)$ and $I$ is a flavour indice  $I = 1,...4$ corresponding to $A_{\mu}, U_{\mu}, V^{\pm}_{\mu}$

For the granular symmetric fields strengths, one gets
\begin{align}
	S^{I}_{\mu \nu} &=\partial_{\mu}A^{I}_{\nu} + \partial_{\nu}A^{I}_{\mu};& S^{\alpha I}_{\alpha}& =2\partial_{\alpha}A^{\alpha I}
\end{align}
\begin{align}
	S^{I}_{0i} &= \partial_{0}A^{I}_{i} + \partial_{i}A^{I}_0, & S^{I}_{ij} &= \partial_{i}A^{I}_{j} + \partial_{j}A^{I}_{i}
\end{align}

The antisymmetric collective fields strengths are written as
\begin{align}
	e_{[\mu \nu]} &= \mathbf{e}_{[IJ]}A^{I}_{ \mu}A^{J}_\nu,& \vec{\mathbf{e}}_{i}&=\mathbf{e}_{[0i]},& \vec{\mathbf{b}}_{i} &= \frac{1}{2}\epsilon_{ijk}\mathbf{e}_{[jk]}
\end{align}
The simetric sector for collective
\begin{align}
	s^{\alpha}_{\alpha IJ} &= \mathbf{e}_{(IJ)}A_{\alpha}^{I}A^{\alpha J}
\end{align}
\begin{eqnarray}
 s_{ij}^{IJ} = \mathbf{e}_{(IJ)}\left(A_{i}^{I}A_{j}^{J}+U_{i}^{J}A_{j}^{J}\right)
\end{eqnarray}
The following Lagrangian is composed by the above fields strengths 
\begin{equation}
	L=L_A+L_S+L_M
\end{equation}
where the antisymmetric sector is
\begin{equation}
    L^{A}_{K} = a_{1}F_{\mu \nu}F^{\mu \nu} + U_{\mu \nu}U^{\mu \nu} + a_{3}V_{\mu \nu}V^{\mu \nu},
\end{equation}
\begin{eqnarray}
    &&L^{A}_{3} = \left(a_{1}F_{\mu \nu} + a_{2}U_{\mu \nu}\right)\left(\mathbf{e}^{[12][\mu \nu]} +\mathbf{e}^{[+-][\mu \nu]}\right)+
    \\
    &&+ a_{3}V^{+}_{\mu \nu}\mathbf{e}^{[12-][\mu \nu]} + a_{3}V^{-}_{\mu \nu}\mathbf{e}^{[12+][\mu \nu]},\nonumber
\end{eqnarray}
\begin{eqnarray}
    L^{A}_{4} = \left(\mathbf{e}^{[12]}_{[\mu \nu]} + \mathbf{e}^{[+-]}_{[\mu\nu]}\right)^{2} + \mathbf{e}^{[12+][\mu \nu]}\mathbf{e}^{[12-][\mu \nu]}.
\end{eqnarray}

The symmetric sector
\begin{eqnarray}
   && L^{S}_{K} = \left(S_{\mu \nu 1} + S_{\mu \nu 2} + g_{\mu \nu}S_{\alpha 1}^{\alpha} + g_{\mu \nu}S_{\alpha 2}^{\alpha} \right)^{2} + \beta_{3}S_{\mu \nu}S^{+}_{\mu \nu}S^{\mu \nu -}  
   \\
   && 4\rho_{3}S^{\alpha +}_{\alpha}S^{\beta-}_{\beta} + \beta_{+}\rho_{-}S^{\alpha+}_{\alpha}S^{\beta-}_{\beta} + \beta_{-}\rho_{+}S^{\alpha-}_{\alpha}S^{\beta +}_{\beta},\nonumber
\end{eqnarray}
\begin{eqnarray}
  && L^{S}_{3} =\left(S_{\mu \nu 1} + S_{\mu \nu 2} + g_{\mu \nu}S_{\alpha 1}^{\alpha} + g_{\mu \nu}S_{\alpha 2}^{\alpha} \right)(\mathbf{e}^{(11)(\mu \nu)} + \mathbf{e}^{(12)}_{(\mu \nu)}+ \nonumber
   \\
    &&\mathbf{e}^{(22)(\mu \nu)}+ \mathbf{e}^{(+-)(\mu \nu)}+ g^{\mu \nu}\mathbf{e}^{(11)\alpha}_{\alpha}  + g^{\mu \nu}\mathbf{e}^{(12)\alpha}_{\alpha}  + g^{\mu \nu}\mathbf{e}^{(22)\alpha}_{\alpha} 
    \\
    &&+ g^{\mu \nu}\mathbf{e}^{(+-)\alpha}_{\alpha}) + (\beta_{+}S^{+}_{\mu \nu} + \rho_{+}g_{\mu \nu}S^{\alpha +}_{\alpha})(\mathbf{e}^{(12-)(\mu \nu)} + g^{\mu \nu}\mathbf{e}^{(12-)\alpha}_\alpha) \nonumber
    \\
    &&+ (\beta_{-}S^{-}_{\mu \nu} + \rho_{-}g_{\mu \nu}S^{\alpha -}_{\alpha})(\mathbf{e}^{(12+)(\mu \nu)} + g^{\mu \nu}\mathbf{e}^{(12+)\alpha}_\alpha),\nonumber
\end{eqnarray}

\begin{eqnarray}
  &&L^{S}_{4} = (\mathbf{e}^{(11)(\mu \nu)} + \mathbf{e}^{(12)}_{(\mu \nu)}+ \mathbf{e}^{(22)(\mu \nu)}+ \mathbf{e}^{(+-)(\mu \nu)}+ g^{\mu \nu}\mathbf{e}^{(11)\alpha}_{\alpha}  \nonumber
  \\
  &&+ g^{\mu \nu}\mathbf{e}^{(12)\alpha}_{\alpha} +
   g^{\mu \nu}\mathbf{e}^{(22)\alpha}_{\alpha} + g^{\mu \nu}\mathbf{e}^{(+-)\alpha}_{\alpha})^2 + (\mathbf{e}^{(12-)(\mu \nu)} + g^{\mu \nu}\mathbf{e}^{(12-)\alpha}_\alpha) 
   \\
   &&(\mathbf{e}^{(12+)(\mu \nu)} + g^{\mu \nu}\mathbf{e}^{(12+)\alpha}_\alpha) + (\mathbf{e}^{(++)}_{(\mu \nu)} + g_{\mu \nu}\mathbf{e}^{(++)\alpha}_\alpha)(\mathbf{e}^{(--)(\mu \nu)} + g^{\mu \nu}\mathbf{e}^{(--)\alpha}_\alpha).\nonumber
\end{eqnarray}

A gauge invariant mass sector is introduced without requiring Higgs mechanism or dynamical symmetry breaking. It gives,
\begin{equation}
    L_{M} = \mathbf{m}^2_{U}U_{\mu}U^{\mu} + \mathbf{m}^{2}_{V}V_{\mu}^{+}V^{\mu -}
\end{equation}
Thus, the quadruplet interrelationships are expressing a four bosons lagrangian. A structure to be investigated

\section{Constitutive equations}

The quadruplet $\{A_{\mu}, U_{\mu}, V_{\mu}^{\pm}\}$  is constituted by fields interrelations. It yields a physics given by constitutive fields equations. They join the corresponding Euler-Lagrange equation, algebraic relationships as $\partial_{\nu}S^{\nu \mu I} = \partial_{\nu}F^{\nu \mu I} + g^{\nu \mu}\partial_{\mu}S^{\alpha}_{\alpha I}$, $\partial_{\nu}\mathbf{e}^{(\nu \mu)} =\frac{1}{2}\mathbf{e}_{IJ}S^{\alpha I}_{\alpha} + \mathbf{e}_{(IJ)}S^{\nu \mu I}A_{\mu}^{J}-\frac{1}{2}\partial^{\mu}\mathbf{e}^{\alpha}_{\alpha} $ and Noether theorem. A whole fields dynamics is derived. For the spin-1 sector, it gives the following non-linear abelian equations.

For $A^{\mu}_{T}$:

The Gauss constitutive law is
\begin{equation}
    \Vec{\nabla}\cdot\{4a_{1}\vec{E}_{A} + a_{2}(\vec{\mathbf{e}}_{AU} + \vec{\mathbf{e}}_{+-})\} = 2a_{3}\mathbf{m}^{2}_{U}\phi_{U} + \rho_{AT} \label{equacao de gauss para A const.}
\end{equation}
with the fields charge density expressed as
\begin{equation}
    \rho_{AT} \equiv \rho_{AT} + a_{3}\rho_{UT} + \rho_{NT}
\end{equation}
where
\begin{eqnarray}
    &&\rho_{AT} \equiv \mathbf{e}_{[12]}[(a_{1}\vec{E}_{A} + E_{U})\cdot\vec{U} + 2(\vec{E}^{+}\cdot\vec{V}^{-} + \vec{E}^{+}\cdot\vec{E}^{-}) + 4(\vec{\mathbf{e}}_{AU} + \vec{\mathbf{e}}_{+-}) \nonumber
    \\
    &&+(\vec{\mathbf{e}}_{AU-}\cdot \vec{V}^{+} + \vec{\mathbf{e}}_{AU+}\cdot \vec{V}^{-})] +2iq_{1}(\frac{\partial \phi^{+}}{\partial t}\phi^{-} + \frac{\partial \vec{V}^{-}}{\partial t}\vec{V}^{+} - \frac{\partial \phi^{+}}{\partial t}\phi^{-}
    \\
    &&-\vec{\nabla}\phi^{+}\cdot\vec{V}^{-} + \frac{\partial \vec{V}^{-}}{\partial t}\cdot\vec{V}^{+} - \vec{\nabla}\cdot\phi^{-}\cdot\vec{V}^{+})-q_{1}^{2}(2\phi^{+}\phi^{-}\phi_{A} + 2\vec{V}^{+}\cdot\vec{V}^{-}\phi_{A} \nonumber
    \\
   && - \vec{V}^{+}\cdot\vec{A}\phi^{-}-\phi^{-}\phi_{A}\phi^{+}-\vec{V}^{-}\cdot\vec{A}\phi^{+})\nonumber
\end{eqnarray}
and

\begin{eqnarray}
    &&\rho_{UT} \equiv \mathbf{e}_{[12]}[(a_{1}\vec{E}_{A} + E_{U})\cdot\vec{A} + 2(\vec{E}^{+}\cdot\vec{V}^{-} + \vec{E}^{+}\cdot\vec{V}^{-}) + 4(\vec{\mathbf{e}}_{AU} + \vec{\mathbf{e}}_{+-}) \nonumber
    \\
    &&+(\vec{\mathbf{e}}_{AU-}\cdot \vec{V}^{+} + \vec{\mathbf{e}}_{AU+}\cdot \vec{V}^{-})] +2iq_{2}(\frac{\partial \phi^{+}}{\partial t}\phi^{-} + \frac{\partial \vec{V}^{-}}{\partial t}\vec{V}^{+} - \frac{\partial \phi^{+}}{\partial t}\phi^{-}
    \\
    &&-\vec{\nabla}\phi^{+}\cdot\vec{V}^{-} + \frac{\partial \vec{V}^{-}}{\partial t}\cdot\vec{V}^{+} - \vec{\nabla}\cdot\phi^{-}\cdot\vec{V}^{+})-q_{2}^{2}(2\phi^{+}\phi^{-}\phi_{U} + 2\vec{V}^{+}\cdot\vec{V}^{-}\phi_{U} \nonumber
    \\
   && - \vec{V}^{+}\cdot\vec{U}\phi^{-}-\phi^{-}\phi_{U}\phi^{+}-\vec{V}^{-}\cdot\vec{U}\phi^{+})
\end{eqnarray}
are derived from Euler-Lagrange equations. Noether theorem provides the conserved electric charge $\rho_{NT}$ expressed as

  \begin{eqnarray}
 \rho_{NT} = iq\{[v_{1}\vec{E}^{-} + v_{2}\vec{e}_{AU-}]\cdot\vec{V}^{+} - [v_{1}\vec{E}^{+} + v_{2}\vec{e}_{AU+}]\cdot\vec{V}^{-}\}
\end{eqnarray}

The Ampère constitutive law is
\begin{eqnarray}
    \vec{\nabla}\times\{4a_{1}\vec{B}_{A} +2a_{2}(\vec{b}_{AU} + \vec{b}_{+-})\}-\frac{\partial}{\partial t}\{4a_{1}\vec{E}_{A} + 2a_{2}(\vec{\mathbf{e}}_{AU} + \vec{\mathbf{e}}_{+-})\} = 2\mathbf{m}^{2}_{U}\vec{U} + \vec{J}_{AT}\label{Eq foton contitutiva}
\end{eqnarray}
with
\begin{equation}
    \vec{J}_{AT} =\vec{j}_{AT} + a_{3}\vec{j}_{UT}+ \vec{j}_{NT}
\end{equation}
where
 \begin{eqnarray}
    &&\vec{j}_{AT} \equiv \mathbf{e}_{[12]}[(a_{1}\vec{E}_{A} + u_{1}\vec{E}_{u} + 2\vec{\mathbf{e}}_{AU} +2\vec{\mathbf{e}}_{+-} )\phi_{U} + (a_{1}\vec{B}_{A} + u_{1}\vec{B}_{u} \nonumber
    \\
    &&+ \vec{b}_{AU} + \vec{b}_{+-})\times\vec{U}-(\vec{E}_{+}+\vec{\mathbf{e}}_{AU+})\phi^{-} -(\vec{B}_{+} + \vec{b}_{AU+})\times\vec{V}^{-}\nonumber
    \\
    &&-(\vec{E}_{-}+\vec{\mathbf{e}}_{AU-})\phi^{+}-(\vec{B}_{-}+\vec{b}_{AU-})\times \vec{V}^{+}]+2iq_{1}(\nabla \phi^{+}\phi^{-} + \nonumber
    \\
    &&\vec{\nabla}\cdot\vec{V}^{+}\cdot\vec{V}^{-} -\frac{\partial \vec{V}^{+}}{\partial t} \phi^{-} + \nabla\phi^{-}\phi^{+} + \vec{\nabla}\cdot \vec{V}^{-}\vec{V}^{+}-\frac{\partial \vec{V}^{-}}{\partial t}\phi^{+} \nonumber
    \\
    &&- \vec{\nabla}\cdot\vec{V}^{+})-q_{1}^{2}(2\phi^{+}\phi^{-}\vec{A} + 2\vec{V}^{+}\cdot\vec{V}^{-}\vec{A} -\phi^{+}\phi_{A}\vec{V}^{-}-\vec{V}^{+}\cdot\vec{A}\vec{V}^{-}\nonumber
    \\
    &&-\phi^{-}\phi_{A}\vec{V}^{-} - (\vec{V}^{-}\cdot\vec{A})\vec{V}^{+})
\end{eqnarray}

and

 \begin{eqnarray}
    &&\vec{j}_{UT} \equiv \mathbf{e}_{[12]}[(a_{1}\vec{E}_{A} + u_{1}\vec{E}_{u} + 2\vec{\mathbf{e}}_{AU} +2\vec{\mathbf{e}}_{+-} )\phi_{A} + (a_{1}\vec{B}_{A} + u_{1}\vec{B}_{u} \nonumber
    \\
    &&+ \vec{b}_{AU} + \vec{b}_{+-})\times\vec{A}-(\vec{E}_{+}+\vec{\mathbf{e}}_{AU+})\phi^{-} -(\vec{B}_{+} + \vec{b}_{AU+})\times\vec{V}^{-}\nonumber
    \\
    &&-(\vec{E}_{-}+\vec{\mathbf{e}}_{AU-})\phi^{+}-(\vec{B}_{-}+\vec{b}_{AU-})\times \vec{V}^{+}]+2iq_{2}(\nabla \phi^{+}\phi^{-} + \nonumber
    \\
    &&\vec{\nabla}\cdot\vec{V}^{+}\cdot\vec{V}^{-} -\frac{\partial \vec{V}^{+}}{\partial t} \phi^{-} + \nabla\phi^{-}\phi^{+} + \vec{\nabla}\cdot \vec{V}^{-}\vec{V}^{+}-\frac{\partial \vec{V}^{-}}{\partial t}\phi^{+} \nonumber
    \\
    &&- \vec{\nabla}\cdot\vec{V}^{+})-q_{2}^{2}(2\phi^{+}\phi^{-}\vec{U} + 2\vec{V}^{+}\cdot\vec{V}^{-}\vec{U} -\phi^{+}\phi_{U}\vec{V}^{-}-\vec{V}^{+}\cdot\vec{U}\vec{V}^{-}\nonumber
    \\
    &&-\phi^{-}\phi_{U}\vec{V}^{-} - \vec{V}^{-}\cdot\vec{U} \vec{V}^{+})
\end{eqnarray}
The Noether current is
\begin{eqnarray}
&&\vec{j}_{NT} = iq\{[v_{1}\vec{B}^{-} + v_{2}\vec{b}_{AU-}]\times\vec{V}^{+} + [[v_{1}\vec{E}^{-} + v_{2}\vec{e}_{AU-}]\phi^{+}]\nonumber
 \\
 &&-[v_{1}\vec{B}^{+} + v_{2}\vec{b}_{AU+}]\times\vec{V}^{-} + [[v_{1}\vec{E}^{+} + v_{2}\vec{e}_{AU+}]\phi^{-}]\}
\end{eqnarray}

The Faraday law for $A^{\mu}_{T}$ is
\begin{eqnarray}
 \vec{\nabla}\times\vec{E}_{A}= -\frac{\partial \vec{B}_{A}}{\partial t}
\end{eqnarray}
with the Gauss law
\begin{eqnarray}
 \vec{\nabla}\cdot\vec{B}_{A} = 0
\end{eqnarray}

The above expressions are expressing a fields EM behavior. In 1820, Oersted discovered a link between electricity and magnetism by observing the current in a wire would move a magnetic needle. A result later explored by Ampère [10] and concluded by Maxwell. However, when one moves from linear to nonlinear EM, new properties emerge. Fields are generating fields and matter. Potential fields physicalities appear. Interactions beyond electric charge are inserted as eqs. (\ref{equacao de gauss para A const.}) and (\ref{Eq foton contitutiva}) are showing.

For $U^{\mu}_{T}$, the constitutive Gauss equation is

\begin{eqnarray}
   \vec{\nabla}\cdot\{4u_{1}\vec{E}_{U} + 2u_{2}\left(\vec{\mathbf{e}}_{AU} + \vec{\mathbf{e}}_{+-}\right)\} -2\mathbf{m}^{2}_{U}\phi_{U} = \rho_{UT}
\end{eqnarray}
where
\begin{equation}
    \rho_{UT} \equiv u_{3}\rho_{AT} + \rho_{UT} + \rho_{NT} 
\end{equation}

The corresponding Ampère law is

\begin{eqnarray}
   &&\vec{\nabla}\times\{4u_{1}\vec{B}_{U} + u_{2}(\vec{b}_{AU} + \vec{b}_{+-})\} - \frac{\partial}{\partial t}\{4u_{2}\vec{E}_{U} + u_{2}(\vec{\mathbf{e}}_{AU} + \vec{\mathbf{e}}_{+-})\} = \nonumber
   \\
   &&=-2\mathbf{m}^{2}_{U}\vec{U} + \vec{J}_{UT}
\end{eqnarray}
where
\begin{eqnarray}
   \vec{J}_{UT} \equiv u_{3}\vec{j}_{AT} + \vec{j}_{UT} + \vec{j}_{NT}
\end{eqnarray}

The Faraday law for $U^{\mu}_{T}$ is given for
\begin{eqnarray}
 \vec{\nabla}\times\vec{E}_{U}=- \frac{\partial \vec{B}_{U}}{\partial t}
\end{eqnarray}
with the Gauss law for
\begin{eqnarray}
 \vec{\nabla}\cdot\vec{B}_{U} = 0
\end{eqnarray}

For $V^{\pm}_{\mu}$, the constitutive Gauss law is

\begin{eqnarray}
   \vec{\nabla}\cdot\{2(v_{1}+\beta_{3})\vec{E}^{\pm} + 2\vec{\mathbf{e}}_{AU\pm}\} - \mathbf{m}^{2}_{V}\phi^{\pm} = \rho^{\pm}_{VT}
\end{eqnarray}

where
\begin{eqnarray}
   &&\rho_{VT}^{\pm} \equiv 2v_{1}\vec{E}^{\pm}\cdot(\vec{A} + \vec{U}) - q_{1}^{2}(\phi^{\pm}\vec{A}\cdot\vec{A} - \vec{V}^{\pm}\cdot\vec{A}\phi_{A})\nonumber
   \\
   &&-q^{2}_{2}(\phi^{\pm}\vec{U}\cdot\vec{U} - \vec{V}^{\pm}\cdot\vec{U}\phi_{U}-4i\mathbf{e}_{[34]}(\vec{\mathbf{e}}_{AU} + \vec{\mathbf{e}}_{+-})\cdot\vec{V}^{\pm} 
   \\
   &&-2iq_{1}(\vec{\nabla}\phi^{\pm}\cdot\vec{A} + \frac{\partial}{\partial t}\vec{V}^{\pm}\cdot\vec{A})-2iq_{2}(\vec{\nabla}\phi^{\pm}\cdot\vec{U} + \frac{\partial}{\partial t}\vec{V}^{\pm}\cdot\vec{U})\nonumber
\end{eqnarray}

The corresponding Ampère law is
\begin{eqnarray}
   &&\vec{\nabla}\times\{2(v_{1} +\beta_{3})\vec{B}^{\pm} + 2v_{2}\vec{b}_{AU\pm}\} - \frac{\partial}{\partial t}\{2(v_{1} +\beta_{3})\vec{E}^{\pm} \nonumber
   \\
   &&+ 2v_{2}\vec{\mathbf{e}}_{AU\pm}\} - \mathbf{m}^{2}_{V}\vec{V}^{\pm} = \vec{J}^{\pm}_{VT}
\end{eqnarray}
where
\begin{eqnarray}
   &&\vec{J}^{\pm}_{VT} \equiv 2v_{1}\vec{B}^{\pm}\times(\vec{A} + \vec{U}) + 2v_{1}\vec{1}\vec{E}^{\pm}\cdot(\phi_{A} + \phi_{U}) -\nonumber
   \\
   &&-q_{1}^{2}(\vec{V}^{\pm}\phi_{A}\phi_{A} + \vec{V}^{\pm}\cdot\vec{A}\vec{A} - \phi^{\pm}\phi_{A}\vec{A} - \vec{V}^{\pm}\vec{A}\cdot\vec{A})-\nonumber
   \\
   &&-q_{2}^{2}(\vec{V}^{\pm}\phi_{U}\phi_{U} + \vec{V}^{\pm}\cdot\vec{U}\vec{U} - \phi^{\pm}\phi_{U}\vec{U} - \vec{V}^{pm}\vec{U}\cdot\vec{U})-\nonumber
   \\
   &&-4i\mathbf{e}_{[34]}[(\vec{b}_{AU} + \vec{b}_{+-})\times\vec{V}^{\pm} + (\vec{\mathbf{e}}_{AU} + \vec{\mathbf{e}}_{+-})\phi^{\pm}]-\nonumber
   \\
   &&-2iq_{1}(\frac{\partial}{\partial t}\vec{V}^{\pm}\phi_{A} - \vec{\nabla}\phi^{\pm}\phi_{A})-2iq_{2}(\frac{\partial}{\partial t}\vec{V}^{\pm}\phi_{U} - \vec{\nabla}\phi^{\pm}\phi_{U})
\end{eqnarray}

Faraday law for $V^{\pm}_{\mu}$
\begin{eqnarray}
  \vec{\nabla}\times\vec{E}^{\pm}+ \frac{\partial \vec{B}^{\pm}}{\partial t} =iq_{1}\{\phi_{A}\vec{B}^{\pm} + 2\vec{A}\cdot\vec{E}^{\pm}\} + iq_{2}\{\phi_{U}\vec{B}^{\pm} + 2\vec{U}\cdot\vec{E}^{\pm}\}\label{Faraday law V+-}
\end{eqnarray}
with the Gauss law 
\begin{eqnarray}
 \vec{\nabla}\cdot\vec{B}^{\pm} = iq_{1}\vec{A}\cdot\vec{B}^{\pm} + iq_{2}\vec{U}\cdot\vec{B}^{\pm}\label{For magnetism V+-} 
\end{eqnarray}
where eqs. (\ref{Faraday law V+-}-\ref{For magnetism V+-}) are introducing sources. 

Thus, the spin-1 sector, introduces nonlinear Faraday lines of force. The couplings between potential fields and fields strengths are the origins for the electromagnetic flux. A nonlinear fields dynamics with conservation laws and fields charges are introduced. Potential fields are explicated at equations of motion and two types of conserved charges beyond electric charge are obtained.  

For the spin-0 longitudinal sector: 

For $A^{\mu}_{L}$:
\begin{eqnarray}
       \partial^{0}\{s_{1}S^{\alpha 1}_{\alpha} + c_{1}\mathbf{e}^{\alpha}_{\alpha}\} - 2t_{1}\mathbf{m}_{U}U^{0} = J^{0}_{AL}
\end{eqnarray}
with

\begin{eqnarray}
      J^{0}_{AL} = j^{0}_{AL} + t_{1}j^{0}_{UL} -(t_{1} +1)J^{0}_{L}
\end{eqnarray}
and

For $U^{\mu}_{L}$:
\begin{equation}
    \partial^{0}\{s_{2}S^{\alpha}_{\alpha 2} +c_{2}s^{\alpha}_{\alpha}\} -2m_{U}U^{\mu}_{L} = J^{0}_{UL}
\end{equation}

For $V_{L}^{\pm}$:

\begin{eqnarray}
      \partial^{0}\{s_{\pm}S^{\alpha \pm} + c_{\pm}s^{(12\pm)\alpha}_{\alpha}\} - \mathbf{m}^{2}_{V}V^{0} = J^{0 \pm}_{VL}
\end{eqnarray}

and 
\begin{eqnarray}
      \partial^{i}\{s_{\pm}S^{\alpha \pm} + c_{\pm}s^{(12\pm)\alpha}_{\alpha}\} - \mathbf{m}^{2}_{V}V^{i} = J^{i \pm}_{VL}
\end{eqnarray}

\section{Collective Bianchi identities}

The quadruplet replies with the usual Bianchi identities. Four granular identities as $\partial_{\mu}F^{I}_{\nu \rho} + \partial_{\rho}F^{I}_{\mu \nu} + \partial_{\nu}F^{I}_{\rho \mu}$ are obtained for $F^{\pm}_{\mu \nu} \equiv \{F_{\mu \nu}, U_{\mu \nu}, V_{\mu \nu}^{\pm}\}$. Nevertheless, due to the quadruplet interrelationships, collective identities are also developed. They are for the pair $\{\vec{e}_{AU}, \vec{b}_{AU}\}:$

\begin{equation}
	\vec{\nabla} \times \vec{\mathbf{e}}_{AU} + \frac{\partial}{\partial t}\vec{b}_{AU} = \mathbf{e}_{[12]}\left(\vec{A} \times \vec{E}_U - \vec{U}\times \vec{E}_A + \phi_A\vec{B}_U - \phi_U\vec{B}_A\right)
\end{equation}
and 
\begin{equation}
	\vec{\nabla}\cdot \vec{b}_{AU} = \mathbf{e}_{[12]}\left(-\vec{A}\cdot \vec{B}_U - \vec{U} \cdot \vec{B}_{A}\right),
\end{equation}

For $\{\vec{\mathbf{e}_{+-}}, \vec{b}_{+-}\}$:
\begin{equation}
	\vec{\nabla} \times \vec{\mathbf{e}_{+-}} + \frac{\partial}{\partial t} \vec{b}_{+-} = -i \mathbf{e}_{[34]}\{\vec{V}^{+} \times \vec{E}_{-} -\vec{V}^{-}\times\vec{E}_{+} -\phi^{+}\vec{B}_{-} + \phi^{-}\vec{B}_{+}\}
\end{equation}
\begin{equation}
	\vec{\nabla}\cdot \vec{b}_{+-} = i\mathbf{e}_{[+-]}\{\vec{V}^{+}\cdot \vec{B}_{-} - \vec{V}^{-}\cdot \vec{B}_{+} \},
\end{equation}

For $\{\vec{e}_{AU+} + \vec{e}_{AU-}, \vec{b}_{AU+}+ \vec{b}_{AU-}\}$: 
\begin{eqnarray}
	&&\vec{\nabla} \times (\vec{b}_{AU+} + \vec{b}_{AU-}) + \frac{\partial}{\partial t}(\vec{\mathbf{e}}_{AU+} +\vec{\mathbf{e}}_{AU-}) = Re\{\mathbf{e}_{[12]}(\vec{A} \times \vec{E}_+\nonumber
	\\
	&&- \vec{V}^+ \times \vec{E}_{A} + \phi_A\vec{B}_+ - \phi_+ \vec{B}_A)\} + Re\{\mathbf{e}_{[12]}\big(\vec{U}\ \times \vec{E}_++\nonumber
	\\
	 &&- \vec{V}^+ \times \vec{E}_{U} + \phi_U\vec{B}_+ - \phi_+ \vec{B}_U)\} + \mathbf{e}_{[12]}\{Req_{1}[\phi^{+}\vec{A}\times\vec{A}\nonumber
	 \\
	 &&+2 \phi_{A}\vec{A}\times\vec{V}^{+}+2\phi_{A}\vec{A}\times\vec{V}^{+}+\phi_{A}\phi_{A}\vec{V}^{+}+2\phi_{A}\phi^{+}\vec{A}\nonumber
	 \\
	&& -\phi_{U}\phi_{A}\phi^{+}\vec{U}-\phi_{U}\phi^{+}\vec{V}^{+}] + Req_{2}[\phi^{+}\vec{U}\times\vec{U}\nonumber
	 \\
	 &&+2 \phi_{U}\vec{U}\times\vec{V}^{+}+2\phi_{U}\vec{U}\times\vec{V}^{+}+\phi_{U}\phi_{U}\vec{V}^{+}+2\phi_{U}\phi^{+}\vec{A}\nonumber
	 \\
	&& -\phi_{U}\phi_{A}\phi^{+}\vec{A}-\phi_{A}\phi^{+}\vec{U}^{+}] \}
\end{eqnarray}

\begin{eqnarray}
	&&\vec{\nabla} \cdot \left(\vec{b}_{+AU} + \vec{b}_{-AU}\right) = Re\mathbf{e}_{[12]}\{ (-\vec{A} \cdot \vec{B}_+ +\vec{V}^{+}\cdot \vec{B}_A) \nonumber
	\\
	&&+(-\vec{U} \cdot \vec{B}_+ +\vec{V}^{+}\cdot \vec{B}_U)\} +Re\mathbf{e}_{[12]}\{q_{1}[2\phi_{A}\vec{A}\cdot\vec{V}^{+}+\phi^{+}\vec{A}\cdot\vec{A}\nonumber
	\\
	&&-\phi_{+}\vec{A}\cdot\vec{U}-\phi_{A}\vec{U}\cdot\vec{V}^{+}] + q_{2}[2\phi_{U}\vec{U}\cdot\vec{V}^{+}+\phi^{+}\vec{U}\cdot\vec{U}-\phi^{+}\vec{U}\cdot\vec{A}-\nonumber
	\\
	&&\phi_{U}\vec{A}\cdot\vec{V}^{+}-\phi_{A}\vec{U}\cdot\vec{V}^{+}] \},
\end{eqnarray}

For symmetric collective fields strengths, with $\{s^{\mu \nu}_{AA} + s^{\mu \nu}_{UU} +s^{\mu \nu}_{AU}, \vec{s}_{AA} + \vec{s}_{UU} + \vec{s}_{AU}\}$:
\begin{eqnarray}
  && \partial^{i}(s^{0j}_{AA} + s^{0j}_{UU} +s^{0j}_{AU}) + \partial^{j}(s^{i0}_{AA} + s^{i0}_{UU} + s^{i0}_{AU}) + \partial^{0}(s^{ij}_{AA} + \partial^{0}s^{ij}_{UU} + s^{ij}_{AU})=\nonumber
  \\
  &&+\mathbf{e}_{(11)}\left\{A^{i}S^{0j}_{A} + A^{j}S^{0i}_{A} + A^{0}S^{ij}\right\} + \mathbf{e}_{(22)}\left\{U^{i}S^{0j}_{U} + U^{j}S^{0i}_{U} + U^{0}S^{ij}\right\}+\nonumber
  \\
  &&+\mathbf{e}_{(12)}\left\{A^{i}S^{0j}_{U} + A^{j}S^{0i}_{U} + A^{0}S^{ij}_{U} + U^{i}S^{0j}_{A} + U^{j}S^{0i}_{A} + U^{0}S^{ij}_{A} \right\}
\end{eqnarray}

\begin{eqnarray}
  	&&\frac{\partial}{\partial t}(\vec{s}_{AA} + \vec{s}_{UU} + \vec{s}_{AU}) = \mathbf{e}_{(11)}\left\{\phi_{A}\cdot \vec{S}_{A} + \vec{A}S_{A}\right\} + \mathbf{e}_{(22)}\left\{\phi_{U}\cdot \vec{S}_{U} + \vec{U}S_{U}\right\}\nonumber
  	  \\
  	&&+\mathbf{e}_{(12)}\left\{\phi_{A} \vec{S}_{U} + \vec{A}S_{U} + \phi_{U} \vec{S}_{A} + \vec{U}S_{A}\right\}
\end{eqnarray}

For 
\begin{eqnarray}
 \frac{\partial}{\partial t}\vec{s}_{+-} = \mathbf{e}_{(34)}Re\left\{\phi_{+}\vec{S}_{+} + \vec{V}^{+}S_{-}\right\} + iq_{1}\{A_{0}V_{i}^{+}V_{j}^{-}+\nonumber
 \\
 V_{0}^{+}A_{i}V_{j}^{-}+ V_{0}^{-}A_{i}V_{j}^{+}\} + iq_{2}\{U_{0}V_{i}^{+}V_{j}^{-} + V_{0}^{+}U_{i}V_{j}^{-}+ V_{0}^{-}U_{i}V_{j}^{+}\},
\end{eqnarray}

\begin{eqnarray}
 	&&\partial^{i}s^{0j}_{+-} + \partial^{j}s^{i0}_{+-} + \partial^{0}s^{ij}_{+-} = \mathbf{e}_{(34)} Re\left\{V^{i+}S^{0j}_{-} + V^{j+}S^{0i}_{-} + A^{0}S^{ij}_-\right\} +\nonumber
 	\\
 	&&+ iq_{1}\{A_{i}V_{j}^{+}V_{k}^{-}+V_{i}^{+}A_{j}V_{k}^{-}+ V_{i}^{-}A_{j}V_{k}^{+}\} + iq_{2}\{U_{i}V_{j}^{+}V_{k}^{-}+\nonumber
 	\\
 	&&V_{i}^{+}U_{j}V_{k}^{-}+ V_{i}^{-}U_{j}V_{k}^{+}\} 
\end{eqnarray}
For $\{s^{\mu \nu}_{AU+}, s^{\mu \nu}_{AU-} \}$:
\begin{eqnarray}
	&&\partial^{i}(s^{0j}_{AU+} + s^{0j}_{AU-} )+ \partial^{j}(s^{i0}_{A+} + s^{i0}_{A-} ) \partial^{0}(s^{ij}_{A+} + s^{ij}_{A-}+ ) =\mathbf{e}_{(12)}Re\{V^{i+}S^{0j}_{A} +\nonumber
	\\
	&&V^{j+}S^{0i}_{A} +V^{0+}S^{ij}_A + A^{i}S^{0j}_{+} + A^{j}S^{0i}_{+} + A^{0}S^{ij}_+ + V^{i+}S^{0j}_{U} + V^{j+}S^{0i}_{U} +
	V^{0+}S^{ij}_U + \nonumber
	\\
	&&U^{i}S^{0j}_{+} + U^{j}S^{0i}_{+} + U^{0}S^{ij}_+\} +q_{1}\{2(A_{i}A_{i}V_{k}^{+} + V_{k}^{+}A_{i}A_{j})+ A_{i}U_{j}V^{+}_{k}\nonumber
	\\
	&& + U_{i}A_{j}V_{k}^{+} + V_{i}^{+}U_{j}A_{k}\} + q_{2}\{2(U_{i}U_{j}V_{k}^{+} + V^{+}_{i}U_{j}A_{k})\nonumber
	\\
	&&+ A_{i}U_{j}V^{+}_{k} + U_{i}A_{j}V_{k}^{+} + V^{+}_{i}U_{j}A_{k}\}
\end{eqnarray}

and
\begin{eqnarray}
	&&\frac{\partial}{\partial t}\left(\vec{s}_{AU+} + \vec{s}_{AU-} \right) = \mathbf{e}_{(34)}[Re\{\phi_{+}\vec{S}_{A} + \vec{V}^{+}S_{A} + \phi_{A}\vec{S}_{+} + \vec{A}S_{+},\nonumber
	\\
	&&\phi_{+}\vec{S}_{U} + \vec{V}^{+}S_{U} + \phi_{U}\vec{S}_{+} + \vec{U}S_{+}\}+Re\{V^{i+}S^{0j}_{U} + V^{j+}S^{0i}_{U} +\nonumber
	\\
	&&V^{0+}S^{ij}_U + U^{i}S^{0j}_{+} + U^{j}S^{0i}_{+} + U^{0}S^{ij}_+\}] +q_{1}\{2(\phi_{A}A_{i}V_{j}^{+} \nonumber
	\\
	&&+ \phi^{+}A_{i}A_{j}) + \phi_{A}U_{i}V^{+}_{j} + \phi_{U}A_{i}V_{j}^{+} + \phi^{+}U_{i}A_{j}\} + q_{2}\{2(\phi_{U}U_{i}V_{j}^{+} \nonumber
	\\
	&&+ \phi^{+}U_{i}A_{j}) + \phi_{A}U_{i}V^{+}_{j} + \phi_{U}A_{i}V_{j}^{+} + \phi^{+}U_{i}A_{j}\}
\end{eqnarray}

\section{Matter from fields associations}

A new Faraday lines physics arises from nonlinearity. Nonlinear fields work as sources for producing matter. Mass and electric charge are derived from fields associations. This leads us to rewrite the previous constitutive equations of motion through London, conglomerate, and current terms.

An origin for mass is obtained from fields conglomerates. Fields condensations as $A_{\mu}A^{\mu}$, $U_{\mu}U^{\mu}$, $A_{\mu}U^{\mu}$ and so on are constituting masses terms. They redefine the $A_{\mu}^{T}$ Gauss law as 
\begin{eqnarray}
    \vec{\nabla}\cdot\{4a_{1}\vec{E}_{A}+ a_{2}(\vec{\mathbf{e}}_{AU}+\vec{\mathbf{e}}_{+-})\}+l^{0}_{AT}+c^{0}_{AT}=2a_{3}\mathbf{m}^{2}_{U}\phi_{U} + \rho_{AT} + \rho_{NT}\label{Massa por campos A}
\end{eqnarray}
where the London term is 
\begin{eqnarray}
    &&l_{AT}^{0} =\{4\mathbf{e}_{[12]}(\phi_{U}\phi_{U} + \vec{U}\cdot\vec{U})-2a_{3}(q_{2}^{2}+q_{1}q_{2})(\phi^{+}\phi^{-} + \vec{V}^{+}\vec{V}^{-})\phi_{A}\}\nonumber
    \\
    &&\{4a_{3}\mathbf{e}_{[12]}(\phi_{A}\phi_{A} + \vec{A}\cdot\vec{A})-2a_{3}(q_{1}^{2}+q_{1}q_{2})(\phi^{+}\phi^{-} + \vec{V}^{+}\vec{V}^{-})\phi_{U}\}
\end{eqnarray}
and the conglomerate association
\begin{eqnarray}
    &&c_{AT}^{0} = 4\mathbf{e}_{[12]}[(\phi_{A}\phi_{U} + \vec{A}\cdot\vec{U})(\phi_{U}+a_{3}\phi_{A}) + 1/2(1+a_{3})(\vec{\mathbf{e}}_{AU-}\cdot\vec{V}^{+}\nonumber
    \\
    &&+\vec{\mathbf{e}}_{AU+}\cdot\vec{V}^{-})]-2q_{1}q_{2}[(\phi^{+}\phi_{U}+ \vec{V}^{+}\cdot\vec{U})\phi^{-}+a_{3}(\phi^{+}\phi_{A}+\vec{V}^{+}\cdot\vec{A})\phi^{-}\nonumber
    \\
    &&+(\phi^{-}\phi_{U}+\vec{V}^{-}\cdot\vec{U})\phi^{+}+a_{3}(\phi^{-}\phi_{A}+\vec{V}^{-}\cdot\vec{A})\phi^{+}]-q_{1}^{2}[(\phi^{+}\phi_{A}+\vec{V}^{+}\cdot\vec{A})\phi^{-}\nonumber
    \\
    &&+(\phi^{-}\phi_{A}+\vec{V}^{-}\cdot\vec{A})\phi^{+}]-2a_{3}q_{2}^{2}[(\phi^{+}\phi_{U}+\vec{V}^{+}\cdot\vec{U})\phi^{-}\nonumber
    \\
    &&+(\phi^{-}\phi_{U}+\vec{V}^{-}\cdot\vec{U})\phi^{+}],
\end{eqnarray}

The above equations are constituting mass condensates terms. Expliciting bilinear scalar terms. The corresponding fields charge density is

\begin{eqnarray}
    &&\rho_{AT} = \mathbf{e}_{[12]}[(a_{1}\vec{E}_{A}+u_{1}\vec{E}_{U})\cdot(\vec{U}+a_{3}\vec{A}) + 1/2(1+a_{3})(\vec{E}^{+}\cdot\vec{V}^{-}+\vec{E}^{-}\cdot\vec{V}^{+})]\nonumber
    \\
    &&-(2iq_{1}+a_{3}q_{2})(\vec{E}^{+}\cdot\vec{V}^{-}+\vec{E}^{-}\cdot\vec{V}^{+}).
\end{eqnarray}

The Amperè law for $A_{\mu}^{T}$ becomes

\begin{eqnarray}
    &&\vec{\nabla}\times[4a_{1}\vec{B}_{A} + a_{2}(\vec{b}_{AU} + \vec{b}_{+-})]-\frac{\partial}{\partial t}[4a_{1}\vec{E}_{A} + 2a_{2}(\vec{\mathbf{e}}_{AU} + \vec{\mathbf{e}}_{+-})] +\vec{l}_{AT} \nonumber
    \\
    &&+ \vec{c}_{AT} = 2a_{3}\mathbf{m}_{U}\vec{U} + \vec{J}_{AT} + \vec{J}_{NT}
\end{eqnarray}
where
\begin{eqnarray}
    &&\vec{l}_{AT} =\{4\mathbf{e}_{[12]}(\phi_{U}\phi_{U} + \vec{U}\cdot\vec{U})-2a_{3}(q_{2}^{2}+q_{1}q_{2})(\phi^{+}\phi^{-} + \vec{V}^{+}\vec{V}^{-})\}\vec{A}\nonumber
    \\
    &&\{4a_{3}\mathbf{e}_{[12]}(\phi_{A}\phi_{A} + \vec{A}\cdot\vec{A})-2a_{3}(q_{1}^{2}+q_{1}q_{2})(\phi^{+}\phi^{-} + \vec{V}^{+}\vec{V}^{-})\}\vec{U},
\end{eqnarray}
and 
\begin{eqnarray}
    &&\vec{c}_{AT} = 4\mathbf{e}_{[12]})[(\phi_{A\phi_{U}} + \vec{A}\cdot\vec{U})(\vec{U} + a_{3}\vec{A})+1/2(1+a_{3})(\vec{b}_{AU-}\times\vec{V}^{+} + \phi^{+}\vec{\mathbf{e}}_{AU-} \nonumber
    \\
    &&+\vec{b}_{AU+}\times\vec{V}^{-} + \phi^{-}\vec{\mathbf{e}}_{AU+} )]-2q_{1}q_{2}[(\phi^{+}\phi_{U} +\vec{V}^{+}\cdot\vec{U})\vec{V}^{-} + (\phi^{-}\phi_{U}+\vec{V}^{-}\cdot\vec{U})\vec{V}^{+}\nonumber
    \\
    &&+(\phi^{+}\phi_{A} + \vec{V}^{-}\cdot\vec{A})\vec{V}^{+}]-q_{1}^{2}[(\phi^{+}\phi_{A}+\vec{V}^{+}\cdot\vec{A})\vec{V}^{-}+(\phi^{-}\phi_{A}+\vec{V}^{-}\cdot\vec{A})\vec{V}^{+}]+\nonumber
    \\
    &&-2a_{3}q_{2}^{2}[(\phi^{+}\phi_{U} + \vec{V}^{+}\cdot\vec{U})\vec{V}^{-} + (\phi^{-}\phi_{A} + \vec{V}^{-}\cdot\vec{A})\vec{V}^{+}],
\end{eqnarray}
and
\begin{eqnarray}
    &&\vec{J}_{AT} = \mathbf{e}_{[12]}[(a_{a}\vec{B}_{A} + u_{1}\vec{B}_{U})\times(a_{3}\vec{A}+\vec{U}) + (a_{1}\vec{E}_{A} + u_{2}\vec{E}_{U})(a_{3}\phi_{A} + \phi_{U})]+\nonumber
    \\
    &&+1/2(1+a_{3} -4iq_{1}+4ia_{3}q_{2} )(\vec{B}^{+}\times\vec{V}^{-} + \vec{B}^{-}\times\vec{V}^{+} + \vec{E}^{+}\phi^{-} + \vec{E}^{-}\phi^{+}),\nonumber
    \\
\end{eqnarray}

Eq. (\ref{Massa por campos A}) is showing the neutral field physics. Differently from usual photons that just transport EM fields the above equations are showing properties of photon-generating EM fields, mass, and neutral couplings.

The condensated Gauss law for $U^{T}_{\mu}$:
\begin{eqnarray}
    \vec{\nabla}\cdot[4u_{1}\vec{E}_{U} + 2u_{2}(\vec{\mathbf{e}}_{AU} + \vec{\mathbf{e}}_{+-})]-2\mathbf{m}^{2}_{U} + l^{0}_{UT} + c^{0}_{UT} = \rho_{U T} + \rho_{NT}
\end{eqnarray}
where
\begin{eqnarray}
  &&l_{UT}^{0} =\{4\mathbf{e}_{[12]}(\phi_{A}\phi_{A} + \vec{A}\cdot\vec{A})-2u_{3}(q_{2}^{2}+q_{1}q_{2})(\phi^{+}\phi^{-} + \vec{V}^{+}\vec{V}^{-})\phi_{U}\}\nonumber
    \\
    &&\{4a_{3}\mathbf{e}_{[12]}(\phi_{U}\phi_{U} + \vec{U}\cdot\vec{U})-2u_{3}(q_{1}^{2}+q_{1}q_{2})(\phi^{+}\phi^{-} + \vec{V}^{+}\vec{V}^{-})\phi_{A}\},
\end{eqnarray}
and
\begin{eqnarray}
  &&c_{UT}^{0} = 4\mathbf{e}_{[12]}[(\phi_{U}\phi_{A}+ \vec{U}\cdot\vec{A})(\phi_{A} + u_{3}\phi_{U}) + 1/2(1+u_{3})(\vec{\mathbf{e}}_{AU-}\cdot\vec{V}^{+}\nonumber
  \\
  &&+ \vec{\mathbf{e}}_{AU-}\cdot\vec{V}^{+})]-2q_{1}q_{2}[(\phi^{+}\phi_{A} + \vec{U}\cdot\vec{A})\phi^{-} + u_{3}(\phi^{+}\phi_{U} + \vec{U}\cdot\vec{U})\phi^{-}\nonumber
  \\
  &&+ (\phi^{-}\phi_{A} + \vec{V}^{-}\cdot\vec{A})\phi^{+}+ (\phi^{-}\phi_{U} + \vec{V}^{-}\cdot\vec{U})\phi^{+} ]-q_{2}^{2}[(\phi^{+}\phi_{U} + \nonumber 
  \\
  +&&\vec{V}^{+}\cdot\vec{U})\phi^{-} + (\phi^{-}\phi_{U}+\vec{V}^{-}\cdot\vec{U})\phi^{+}]-2u_{3}q_{1}^{2}[(\phi_{+}\phi_{A} + \vec{V}^{+}\vec{A})\phi^{-}\nonumber
  \\
  &&+ (\phi_{-}\phi_{A} + \vec{V}^{-}\vec{A})\phi^{+}],
\end{eqnarray}
and
\begin{eqnarray}
  &&\rho_{UT} = \mathbf{e}_{[12]}[(a_{1}\vec{E}_{A} + u_{2}\vec{E}_{U})\cdot(\vec{A} + u_{3}\vec{U}) + 1/2(1+a_{3})(\vec{E}^{+}\cdot\vec{V}^{-} + \vec{E}^{-}\cdot\vec{V}^{+})]+\nonumber
  \\
  &&-2i(u_{3}q_{1}+ q_{2})(\vec{E}^{+}\cdot\vec{V}^{-} + \vec{E}^{-}\cdot\vec{V}^{+}),
\end{eqnarray}

Ampère law for $U_{\mu}^{T}$:
\begin{eqnarray}
  &&\vec{\nabla}\times[4u_{1}\vec{B}_{U} + u_{2}(\vec{b}_{AU} + \vec{b}_{+-})]-\frac{\partial}{\partial t}[u_{1}\vec{E}_{A} + 2u_{2}(\vec{\mathbf{e}}_{AU} + \vec{\mathbf{e}}_{+-})]-2\mathbf{m}^{2}_{U}\vec{U}+ \vec{l}_{UT} + \nonumber
  \\
  &&\vec{c}_{UT} = \vec{J}_{UT} + \vec{J}_{NT}
\end{eqnarray}

where
\begin{eqnarray}
   &&\vec{l}_{UT} =\{4\mathbf{e}_{[12]}(\phi_{A}\phi_{A} + \vec{A}\cdot\vec{A})-2u_{3}(q_{2}^{2}+q_{1}q_{2})(\phi^{+}\phi^{-} + \vec{V}^{+}\vec{V}^{-})\}\vec{U}\nonumber
    \\
    &&\{4a_{3}\mathbf{e}_{[12]}(\phi_{U}\phi_{U} + \vec{U}\cdot\vec{U})-2u_{3}(q_{1}^{2}+q_{1}q_{2})(\phi^{+}\phi^{-} + \vec{V}^{+}\vec{V}^{-})\}\vec{A},
\end{eqnarray}
and
\begin{eqnarray}
   &&\vec{c}_{UT} = 4\mathbf{e}_{[12]}[(\phi_{A}\phi_{U}+ \vec{A}\cdot\vec{U})(u_{3}\vec{U} + \vec{A})+1/2(1+u_{3})(\vec{b}_{AU-}\times\vec{V}^{+}\nonumber 
   \\
   &&+ \vec{b}_{AU+}\times\vec{V}^{-} + \phi^{+}\vec{\mathbf{e}}_{AU-} + \phi^{-}\mathbf{e}_{AU+})]-2q_{1}q_{2}[(\phi^{+}\phi_{U} + \vec{V}^{+}\cdot\vec{U})\vec{V}^{-} \nonumber
   \\
   &&+ (\phi^{-}\phi_{U} + \vec{V}^{-}\cdot\vec{U})\vec{V}^{+}]-u_{3}q_{1}^{2}[(\phi^{+}\phi_{A}+ \vec{V}^{+}\cdot\vec{A})\vec{V}^{-}+(\phi^{-}\phi_{A}+ \vec{V}^{-}\cdot\vec{A})\vec{V}^{-}]  \nonumber
   \\
   &&2q_{2}^{2}[(\phi^{+}\phi_{U}+ \vec{V}^{+}\cdot\vec{U})\vec{V}^{-}+(\phi^{-}\phi_{U}+ \vec{V}^{-}\cdot\vec{U})\vec{V}^{-}],
\end{eqnarray}
and
\begin{eqnarray}
   &&\vec{J}_{UT} = \mathbf{e}_{[12]}[(a_{1}\vec{B}_{A} + u_{1}\vec{B}_{U})\times(\vec{A}+u_{3}\vec{U})+ (a_{1}\vec{E}_{A} +u_{2}\vec{B}_{U})\cdot(u_{3}\phi_{U} + \phi_{A}) \nonumber
   \\
   &&+ 1/2(1+u_{3})(\vec{B}^{+}\times\vec{V}^{-} + \vec{B}^{-}\times\vec{V}^{+} + \vec{E}^{+}\phi^{-} + \vec{E}^{-}\phi^{+})]\nonumber
   \\
   &&-2i(q_{1} +u_{3}q_{2})(\vec{B}^{+}\times\vec{V}^{-} + \vec{B}^{-}\times\vec{V}^{+} + \vec{E}^{+}\phi^{-} + \vec{E}^{-}\phi^{+}).
\end{eqnarray}

The condensate Gauss law for $V_{\mu}^{\pm}$ is
\begin{eqnarray}
   \vec{\nabla}\cdot\{2(v_{1}+\beta_{3})\vec{E}^{+} + 2v_{2}\vec{e}_{AU\pm}\}-\mathbf{m}^{2}_{V}\phi^{\pm} + l_{VT}^{0 \pm} + c_{VT}^{0 \pm} = \rho^{VT}
\end{eqnarray}
where
\begin{eqnarray}
   &&l_{VT}^{0\pm} = -4i\mathbf{e}_{[34]}\{(\phi^{\pm}\phi^{\pm} + \vec{V}^{\pm}\cdot\vec{V}^{\pm})\phi^{\mp} + \vec{e}_{AU}\cdot\vec{V}^{\pm}\}\nonumber
   \\
   &&-q_{1}^{2}\{\phi_{A}\phi_{A} + \vec{A}\cdot\vec{A}\}\phi^{\pm} - q_{2}^{2}\{\phi_{U}\phi_{U} + \vec{U}\cdot\vec{U}\}\phi^{\pm},
\end{eqnarray}

\begin{eqnarray}
   &&c^{0\pm}_{VT} = -4i\mathbf{e}_{[34]}\{(\phi^{\pm}\phi^{\mp} + \vec{V}^{\pm}\cdot\vec{V}^{\mp})\phi^{\pm} + \mathbf{e}_{AU}\cdot\vec{V}^{\pm}\}\nonumber
   \\
   &&-2q_{1}q_{2}\{(\phi_{A}\phi_{U} + \vec{A}\cdot\vec{U})\phi^{\pm} - (\phi^{\pm}\phi_{U} + \vec{V}^{\pm}\cdot\vec{U})\phi_{A}\} +\nonumber
   \\
   &&q_{1}^{2}\{\phi^{\pm}\phi_{A} + \vec{V}^{\pm}\cdot\vec{A}\}\phi_{A}+q_{2}^{2}\{\phi^{\pm}\phi_{U} + \vec{V}^{\pm}\cdot\vec{U}\}\phi_{U},
   \end{eqnarray}
   
   \begin{eqnarray}
    \rho^{\pm}_{VT} =  2v_{1}\vec{E}^{\pm}\cdot(\vec{A} +  \vec{U}) - 2iq_{1}\vec{E}^{\pm}\cdot \vec{A} - 2iq_{2}\vec{E}^{\pm}\cdot \vec{U}
   \end{eqnarray}
   The associated $V_{\mu}^{\pm}$ Ampère law is
 \begin{eqnarray}
  &&\vec{\nabla}\times\{2(v_{1}  + \beta_{3})\vec{B}^{\pm} + 2\vec{b}_{AU\pm}\} - \frac{\partial}{\partial t}\{2(v_{1}+\beta_{3})\vec{E}^{\pm} + 2\vec{e}_{AU\pm}\} -\mathbf{m}^{2}_{V}\vec{V}^{\pm} \nonumber
  \\
  &&+ \vec{l}_{VT}^{\pm} + \vec{c}^{\pm}_{VT} = \vec{J}^{\pm}_{VT}
 \end{eqnarray}
 where
 \begin{eqnarray}
  &&\vec{l}_{VT}^{\pm} =  -4i\mathbf{e}_{[34]}\{(\phi^{\pm}\phi^{\pm} + \vec{V}^{\pm}\cdot\vec{V}^{\pm})\vec{V}^{\mp} + \vec{e}_{AU}\phi^{\pm} + \vec{b}_{AU}\times\vec{V}\}\nonumber
   \\
   &&-q_{1}^{2}\{\phi_{A}\phi_{A} + \vec{A}\cdot\vec{A}\}\vec{V}^{\pm} - q_{2}^{2}\{\phi_{U}\phi_{U} + \vec{U}\cdot\vec{U}\}\vec{V }^{\pm},
 \end{eqnarray}
 and
 \begin{eqnarray}
  &&\vec{c}^{\pm}_{VT} = -4i\mathbf{e}_{[34]}\{(\phi^{\pm}\phi^{\mp} + \vec{V}^{\pm}\cdot\vec{V}^{\mp})\vec{V}^{\pm} + \mathbf{e}_{AU}\phi^{\pm} + \vec{b}_{AU}\times\vec{V}^{\pm}\}\nonumber
   \\
   &&-2q_{1}q_{2}\{(\phi_{A}\phi_{U} + \vec{A}\cdot\vec{U})\vec{V}^{\pm} - (\phi^{\pm}\phi_{U} + \vec{V}^{\pm}\cdot\vec{U})\vec{A}\} +\nonumber
   \\
   &&q_{1}^{2}\{\phi^{\pm}\phi_{A} + \vec{V}^{\pm}\cdot\vec{A}\}\vec{A}+q_{2}^{2}\{\phi^{\pm}\phi_{U} + \vec{V}^{\pm}\cdot\vec{U}\}\vec{U},
 \end{eqnarray}
 
 and
 
  \begin{eqnarray}
    &&\vec{J}^{\pm}_{VT} =  2v_{1}[\vec{B}^{\pm}\times(\vec{A} +  \vec{U}) + \vec{E}^{\pm}(\phi_{A} + \phi_{U})]- 2iq_{1}(\vec{B}^{\pm}\times \vec{A} + \vec{E}^{\pm}\phi_{a})\nonumber
    \\
    &&- 2iq_{2}(\vec{B}^{\pm}\times \vec{U} + \vec{E}^{\pm}\phi_{A})
   \end{eqnarray}
   
   For longitudinal sector
For $A^{\mu}_{L}$
\begin{eqnarray}
  \partial^{0}\{s_{1}S^{\alpha 1 }_{\alpha} + c_{1}\mathbf{e}^{\alpha}_{\alpha}\}+l_{AL}^{0} +c_{AL}^{0} =2t_{1}\mathbf{m}^{2}_{U}U^{0}+J^{0}_{AL} + (t_{1}+1)J^{0}_{NL}
\end{eqnarray}

\begin{eqnarray}
 && l_{ATL}^{0} = 4\mathbf{e}_{(11)}\mathbf{e}_{(11)}[\mathbf{e}^{(11)(00)}A_{0}+\mathbf{e}^{(11)(0i)}A_{i} + 4 (\mathbf{e}^{(11)\alpha}_{\alpha}+\mathbf{e}^{(12)\alpha}_{\alpha})A^{0}]\nonumber
 \\
 &&+2\mathbf{e}_{(12)}(2\mathbf{e}^{(12)(00)} + \mathbf{e}^{(22)(00)})U_{0} + (2\mathbf{e}^{(12)(0i)} + \mathbf{e}^{(22)(0i)})U_{i} + \nonumber
 \\
 &&+4(\mathbf{e}^{(12)(00)}A_{0} + \mathbf{e}^{(12)(0i)}A_{i}) + \mathbf{e}^{(11)\alpha}_{\alpha}A^{0} + \mathbf{e}^{(22)\alpha}_{\alpha}U^{0})+\nonumber
 \\
 &&+4t_{1}[(\mathbf{e}^{(22)\alpha}_{\alpha} + \mathbf{e}^{(12)\alpha}_{\alpha})U^{0} + \mathbf{e}^{(22)(00)}U_{0} + \mathbf{e}^{(22)(0i)}U_{i}]  +\nonumber
 \\
 &&+2t_{1}\mathbf{e}_{(12)}[2\mathbf{e}^{(11)(00)}A_{0} + \mathbf{e}^{(11)(0i)}A_{i} + \mathbf{e}^{(12)(00)}U_{0} + \mathbf{e}^{(12)(0i)}U_{i}+\nonumber  
 \\
 &&+ (\mathbf{e}^{(22)\alpha}_{\alpha} + \mathbf{e}^{(12)\alpha}_{\alpha})U^{0}]-q_{1}^{2}(\phi^{+}\phi^{-} + \vec{V}^{+}\cdot\vec{V}^{-})A^{0}\nonumber
 \\
 &&+2q_{2}^{2}t_{2}(\phi^{+}\phi^{-}+\vec{V}^{+}\cdot\vec{V}^{-})U^{0} -2q_{1}q_{2}(\phi^{+}\phi^{-} + \vec{V}^{+}\cdot\vec{V}^{-})(A^{0}+ U^{0}) \nonumber
 \\
\end{eqnarray}
and
\begin{eqnarray}
 &&c^{0}_{AL} = 2\mathbf{e}_{(12)}[\mathbf{e}^{(\alpha)}_{\alpha}(U^{0}+A^{0}) + (\mathbf{e}^{(11)(00)} + \mathbf{e}^{(+-)(00)})U_{0} +\nonumber
 \\
 &&+(\mathbf{e}^{(11)(0i)} + \mathbf{e}^{(+-)(0i)})U_{i} + \mathbf{e}^{(+-)\alpha}_{\alpha}U^{0}] + \mathbf{e}_{(11)}(\mathbf{e}^{(12)(00)} \nonumber
 \\
 &&+ \mathbf{e}^{(+-)(00)} )A_{0} + (\mathbf{e}^{(22)(0i)} + \mathbf{e}^{(+-)(0i)})A_{i} + 8\mathbf{e}^{(+-)\alpha}_{\alpha}A^{0}]\nonumber
 \\
 &&+2\mathbf{e}_{(34)}(\mathbf{e}^{(12+)(00)}V_{0}^{-} + \mathbf{e}^{(12-)(00)}V_{0}^{+} + \mathbf{e}^{(12+)(0i)}V_{i}^{-} + \mathbf{e}^{(12-)(0i)}V_{i}^{+}\nonumber
 \\
 &&+ \mathbf{e}^{(12+)\alpha}_{\alpha}V^{0 -} + \mathbf{e}^{(12-)\alpha}_{\alpha}V^{0+}) + 4t_{1}\mathbf{e}_{(22)}[(\mathbf{e}^{(11)(00)} + \mathbf{e}^{(+-)(00)})U_{0} \nonumber
 \\
 &&+ (\mathbf{e}^{(11)(0i)} + \mathbf{e}^{(+-)(0i)})A_{i} + \mathbf{e}^{(+-)\alpha}_{\alpha}U^{i}] + 2t_{1}\mathbf{e}_{12}[\mathbf{e}^{(12)\alpha}_{\alpha}A^{0} + \mathbf{e}^{(12)\alpha}_{\alpha}U^{0}]\nonumber
 \\
 &&+2t_{1}\mathbf{e}_{(34)}(\mathbf{e}^{(12+)(00)}V_{0}^{-} + \mathbf{e}^{(12-)(00)}V_{0}^{+} + \mathbf{e}^{(12+)(0i)}V_{i}^{-} + \mathbf{e}^{(12-)(0i)}V_{i}^{+} \nonumber
 \\
 &&+ \mathbf{e}^{(12+)\alpha}_{\alpha}V^{0 -} + \mathbf{e}^{(12-)\alpha}_{\alpha}V^{0+}) + 2t_{1}q_{2}^{2}[(V_{0}^{+}A^{0} + V_{i}^{+}A^{i})V^{0 -} +\nonumber
 \\
 &&+(V_{0}^{-}A^{0} + V_{i}^{-}A^{i})V^{0 +}]-2q_{1}q_{2}[t_{1}(V_{0}^{+}A^{0} + V_{i}^{+}A^{i})V^{0 -})V^{0-}\nonumber
 \\
 &&+ (V_{0}^{-}A^{0} + V_{i}^{-}A^{i})V^{0 +} + (V_{0}^{+}U^{0} + V_{i}^{+}U^{i})V^{0 -} + (V_{0}^{-}U^{0} + V_{i}^{-}U^{i})V^{0 +}]\nonumber
 \\
\end{eqnarray}
and
\begin{eqnarray}
 &&J^{0}_{AL} = 2(\beta_{1}S^{001} + \beta_{2}S^{002})[\mathbf{e}_{(11)}(1+t_{1})A_{0} + \mathbf{e}_{(11)}U_{0}+ t_{1}\mathbf{e}_{(22)}U_{0}]\nonumber
 \\
 &&+2(\beta_{1}S^{0i1}+ \beta_{2}S^{0i2})[\mathbf{e}_{(11)}(1+t_{1})A_{i} + \mathbf{e}_{(12)}U_{i} + t_{1}\mathbf{e}_{(12)}U_{i}]+\nonumber
 \\
 &&2[(\beta_{1} +
  5\rho_{1})S^{\alpha 1}_{\alpha} + (\beta_{2} + 5\rho_{2})S^{\alpha 2 }_{\alpha}][\mathbf{e}_{(11)}(1+t_{1})A^{0} + \mathbf{e}_{12}(1+t_{1})U^{0} +\nonumber
  \\
  &&+\mathbf{e}_{(22}U^{0}] + \mathbf{e}_{(34)}(1+t_{1})[\beta_{1}(S^{00+}V_{0}^{-} + S^{0i+}V_{i}^{+}) + \beta_{-}(S^{00-}V_{0}^{+} +\nonumber
  \\
  &&+S^{0i-}V_{i}^{+}) + \rho_{+}S^{\alpha +}_{\alpha}V^{0-} + S^{\alpha - }_{\alpha}V^{0+}] 
\end{eqnarray}
and
\begin{eqnarray}
  \partial^{i}\{s_{1}S^{\alpha 1 }_{\alpha} + c_{1}\mathbf{e}^{\alpha}_{\alpha}\}+l_{AL}^{i} +c_{AL}^{i} =2t_{1}\mathbf{m}^{2}_{U}U^{i}+J^{i}_{AL} + (t_{1}+1)J^{i}_{NL}
\end{eqnarray}

For $U^{\mu}_{L}$
\begin{eqnarray}
  \partial^{0}\{s_{2}S^{\alpha 2 }_{\alpha} + c_{2}\mathbf{e}^{\alpha}_{\alpha}\}+l_{UL}^{0} +c_{UL}^{0} -2\mathbf{m}^{2}_{U}U^{0}=J^{0}_{UL} + (t_{2}+1)J^{0}_{NL}
\end{eqnarray}

And
\begin{eqnarray}
  \partial^{i}\{s_{2}S^{\alpha 2 }_{\alpha} + c_{2}\mathbf{e}^{\alpha}_{\alpha}\} -2\mathbf{m}^{2}_{U}U^{i}+l_{AL}^{i} +c_{AL}^{i} = J^{i}_{AL} + (t_{1}+1)J^{i}_{NL}
\end{eqnarray}

For $V^{\pm}_{L}$
\begin{eqnarray}
 \partial^{0}\{s_{\pm}S^{\alpha\pm}_{\alpha} + c_{1}\mathbf{e}^{(12\pm)\alpha}_{\alpha}\} +l_{VL}^{0\pm} + c^{0}_{VL} -\mathbf{m}^{2}_{V}V^{0\pm} = J^{0\pm}_{VL}
\end{eqnarray}
\begin{eqnarray}
 \partial^{i}\{s_{\pm}S^{\alpha\pm}_{\alpha} + c_{1}\mathbf{e}^{(12\pm)\alpha}_{\alpha}\} +l_{VL}^{i\pm} + c^{i}_{VL} -\mathbf{m}^{2}_{V}V^{i\pm} = J^{i\pm}_{VL}
\end{eqnarray}

The above expressions are showing dynamically on EM fields associations. Next step will be to understand show these fields products are transformed into a condensation mechanism.
\section{Conclusion}

Faraday anteceded Einstein on Newton extension. Introduced a physics depending on fields that Newton epoch did not recognize. EM would be guided by variables diverse from mechanics. Nature provides not only matter as fields. Maxwell equations projected fields as variables and matter as sources. 

Faraday used field lines to signal that there was something. His law of induction was definitive. The lines of force became tangible effects. EM is redefined as a flow of fields. Maxwell states it in terms of divergent and rotational operators. The question would be whether these fields formed a substance of their own. The limitation in Faraday is the EM fields depending on electrical charge. Therefore, they would be consequential.

History grows by, and, in current days, physics is based on fields. The fields play the role of the variables of quantum field theory. Faraday fields led to a viewpoint where quantum fields precede particles. The next stage is to explore nonlinear fields dynamics. Different subjects such as plasma, astrophysics, and condensated matter are requiring nonlinear extensions. EM literature provides 15 nonlinear models [11]. Most of them are effective theories, as Born-Infeld and Euler-Heisenberg. In this work, a nonlinear abelian gauge theory is considered. The simplest nonlinear fields dynamics without requiring non-linear Yang-Mills is achieved by introducing different potential fields rotating under the same group [12].

A four bosons EM is proposed. A fundamental nonlinear electromagnetism based on the generic electric charge $\{+,0,-\}$ transmitted by a quadruplet $\{A_{\mu}, U_{\mu}, V_{\mu}^{\pm}\}$ connected under an abelian gauge symmetry, eqs. (2.1-2.4). This association of new EM provides physicalities. Fields equations are providing potential fields with dynamics, conserved charges and couplings. A fields charge physics is derived. Non linear fields are emitting Faraday lines of force. 
A new texture appears through the proposal of abelian non-linear electrodynamics. Its novelty compared to Faraday lies in the existence of potential fields acting as their own sources Equations of motion show fields charges as a function of fields. Gauge invariant, potential fields conglomerates emerge. The photonic field observed in interference and diffraction, introduced by hand in quantum mechanics and detected in the Ahranov-Bohm and Ahranov-Casher experiments, gains presence by fundamental principles.

An EM model with nonlinear abelian potential fields is proposed. Potential fields are no more subsidiary as in Maxwell. They work as their own sources and interact with granular and collective EM fields strengths. Their explicit participation in fields equations is consistent with the Faraday experiment in 1841 when light polarization was shifted by a magnetic field [13], and according to Aharonov-Bohm and Aharonov-Casher experiments [14]. Originating mass through London and conglomerated terms. Electric charge is no more the only conserved charge and coupling constant. Neutral couplings with granular and collective fields strengths are also ruling the model. 

Four bosons EM leads to a physics based on the potential fields physicality. Instead of each point in space is associated with E-B, that is, configured in 6 numbers, a new determination is observed. The EM origin is given by the cluster ${A_{\mu}, U_{\mu}, V_{\mu}^{\pm}} $where each point P is described by 4x4 numbers. Therefore, these 16 numbers define the EM field substance.

Thus, preserving the two Maxwell postulates, which are light invariance and electric charge conservation, the four bosons electromagnetism redefines what electromagnetic energy is. Potential fields conglomerates, granular and collective fields strengths are extending the EM observable beyond the usual pair $\vec{E}$-$\vec{B}$. New sectors are ascribed beyond Maxwell.  Analyzing the corresponding energy expression in [15], one notices other terms than the usual Maxwell $\vec{E}^{2}+\vec{B}^{2}$. Nonlinearity, neutral, weak interaction, spintronics, and photonics regimes are introduced [16]. 

There is an extended EM energy to be understood. Missing energy is always a present theme in physics. In 1927, Ellis and Wooster by studying the average distribution of $\beta$-radium understood that something was missing [17]. Later Pauli identified as neutrinos. Actually, physics discusses dark energy and matter. Something again is absent. Various hypotheses have been considered. Perhaps one should include an EM energy extension.

Concluding, there is a deeper relationship between matter and fields to be investigated by a nonlinear Faraday lines of force. A physics that antecedes matter with dynamics, conservation and charge expressed by fields. Consider particles production by electric, magnetic, and gravitational fields [18]. A subject that has been studied since 1931 by Sauter, Breit-Wheeler in (1934), Schwinger (1951), Sexl (1969). There is a new Faraday field physics to be included in electromagnetic theory.


\begin{thebibliography}{30}

\bibitem{FARADAY}
M. Faraday, Experimenatal researches in electricity. Philosophical transactions of the Royal Society of London, n. 122 p. 125-162, (1832); Experimental Researches in Electricity, edited by Taylor and T. Francis, vols I (1839), II (1844), and Vol. III (1855); On the various forces in Nature (1873).

\bibitem{RUTHERFORD}
E. Rutherford, LXXIX. The scattering of $\alpha$ and $\beta$ particles by matter and the structure of the atom. The London, Edinburgh, and Dublin Philosophical Magazine and Journal of Science, v. 21, n. 125, p. 669-688, (1911).

\bibitem{Bohr}
N. Bohr, XXXVII. On the constitution of atoms and molecules." The London, Edinburgh, and Dublin Philosophical Magazine and Journal of Science 26.153, 476-502, (1913).

\bibitem{MAGNETIC FIELD}
F. Govoni et al, A radio ridge connecting two galaxy clusters in a filament of the cosmic web, Science 364, 981 (2019); R. J. van Weeren, Fast magnetic field amplification in distant galaxy clusters, Nature Astronomy 5, 268 (2021).


\bibitem{MAGNETIC SCALAR 109T}
Ling-Da Kong et al, Insight-HXMT Discovery of the Highest-energy CRSF from the First Galactic Ultraluminous X-Ray Pulsar Swift J0243.6+6124, The Astrophysical Journal Letters, (2022). DOI: https://doi.org/10.3847/2041-8213/ac7711. 

\bibitem{Sauter SCHWINGER}
F. Sauter, "Über das Verhalten eines Elektrons im homogen elektrischen Feld nach der relativistischen Theorie Diracs", Zeitshrift für Physik, Springer Science and Business Media LLC, 69(11-12): 742-764, (1931); J. Schwinger, On Gauge Invariance and Vacuum Polarization, Physical Review, American Physical Society (APS). 82(5): 664-679, (1951).

\bibitem{Maxwell}
J.C. Maxwell, On Faraday’s lines of force, Trans. Cambridge Philosoph. Soc., Vol. X,(Part I), 155 (1855); On physical lines of force, Philos. Mag., vol. XXI, 90, (1861);A dynamical theory of the electromagnectic field, Proceedings of the Royal Society of London XIII, 531, (1864), Phylos. Trans. in Royal Soc. of London 155, vol. 459,(1865);  A Treatise on Electricity and Magnetism, Oxford:  Clarendon Press Series,(1873).

\bibitem{FOUR-BOSONS}
R. Doria, A new model for a non-linear electromagnetic model with selfinteracting photons, JAP, 7(3), 1840-1896 (2015);J. Chauca, R. Doria, I. Soares, Four Bosons electromagnetism, JAP, Vol 10 no1 2605 (2015); J. Chauca, R. Doria, I. Soares, Electric Charge transmission through Four Bosons, JAP, 13(i): 4535-4552, (2017); R. Doria, and I. Soares, Four Bosons EM Conservation Laws, JAP. 19. 40-92. (2021). R. Doria, and I. Soares, Spin-Valued Four Bosons Electrodynamics, JAP, 19. 93-133 (2021).

\bibitem{Ultimos trabalhos}
R. Doria and L.S. Mendes,On the electromagnetic symmetry producing fields charges at four bosons association, submitted paper (2022).

\bibitem{Ampère}
A. Ampère, Théorie des phénomenes électro-dynamiques uniquement déduite de l'experience, A. Hermann, Librarie Scientifique (1826).

\bibitem{15 models}
J.A. Helayel-Neto, \url{https://ava.professorglobal.com.br/mod/resource/view.php?id=7907}, (2022).

\bibitem{Yang-Mills abelian}
For a Kaluza-Klein origin see: R.M. Doria and C. Pombo, Two Potentials, One Gauge Group: A Possible Geometrical Interpretation II. Nuov. Cim., (1986); C.M. Doria, R.M. Doria, J. A. Helay{\"e}l-Neto, A Kaluza-Klein Interpretation of an Extended Gauge Theory, Rev. Bras. Fis., v.17, p.351-359, (1987). For supersymmetric origin see: N. Chair, J. A. Helay{\"e}l-Neto and A. William Smith, A less constrained (2, 0) super-Yang-Mills model, Phys. Lett. B,233,173, (1989); S.A. Dias, R.M. Doria, J. L. Matheus Valle, A constraint analysis for an N = 112, D 2 supersymmetric model, Rev. Bras. Fis., 21, 1, (1991); C.M. Doria, R.M. Doria, and F.A.B.R. Carvalho, A superspace origin for an extended Gauge model, Acta Physica Hungarica, v.73, p.51-58, (1993); C.A.S. Almeida, R. M. Doria, Information on the Gauge principle from a (2,0) Supersymmetric Gauge Model Rev. Bras. Fis.,21,3, (1991). - For fibre bundle origin: C. Doria, R.M. Doria, J.A. Helay{\"e}l, A Fiber Bundle Treatment to a class of Extended Gauge Models, Communications in Theoretical Physics, v.17,p.505-508, (1992). For o-model origin see: R. M. Doria, J. A. Helay{\"e}l-Neto and S. Mokhtari, An Extended Gauge Model as a Possible Origin for Nonlinear $\sigma$-Models Europhys. Lett., 16(1),23 (1991); R.M. Doria and J.A. Helay{\"e}l-Neto, A two-gauge field induced CP-model, Rev. Bras. Fis., Vol. 19, $n^1$ (1989); Almeida, C. A., Chauca, J. . and Doria, R. A less-constrained (2,0) super-yang-mills model the coupling to non-linear $\sigma$-models. journal of advances in physics, 20, 209–114, (2022). https://doi.org/10.24297/jap.v20i.9134.

\bibitem{Faraday light polarization}
M. Faraday, XIX. On some supposed forms of lightning: To the editors of the Philosophical Magazine and Journal. The London, Edinburgh, and Dublin Philosophical Magazine and Journal of Science, v. 19, n. 122, p. 104-106, (1841).

\bibitem{Ahranov-Bohm and Ahranov casher}
Y. Aharonov, D. Bohm Significance of electromagnetic potentials in the quantum theory. Physical Review, v. 115, n. 3, p. 485, (1959); Y. Aharonov, A. Casher,  Topological quantum effects for neutral particles. Physical Review Letters, v. 53, n. 4, p. 319, (1984).

\bibitem{Conservations laws}
Doria, and I. Soares, Four Bosons EM Conservation Laws, JAP. 19. 40-92. (2021).

\bibitem{Maxwell to photonics}
M. Couto and R. Doria, Maxwell to photonics, submetted paper, (2022).

\bibitem{Ellis and Wooster}
C. D. Ellis e W. A. Wooster, The average distribution of disintegration of Radium-E, Proc. Roy. Soc. London, PRSL A117, 119 (1927).

\bibitem{Produção de particulas gravitaçaõ}
R. U. Sexl, H.K. Urbantke,  Production of particles by gravitational fields. Physical Review, v. 179, n. 5, p. 1247, (1969).
\end{thebibliography}
\end{document}